\documentclass[10pt]{article}
\usepackage{amssymb}
\usepackage{amsmath}
\usepackage{graphicx}
\usepackage{fancyhdr}
\usepackage{cite}

\begin{document}


\title{Comment on electron-impact excitation cross-section measurements for He-like Xenon}
\author{J.-C. Pain and M. Comet\\
CEA, DAM, DIF, F-91297 Arpajon, France\\
\\
C.J. Fontes\\
Los Alamos National Laboratory, Los Alamos, New Mexico 87545, USA}
\maketitle

\begin{abstract}
We discuss lower-than-predicted collisional-excitation cross-sections for helium-like xenon measured at an Electron Beam Ion Trap facility. In a review paper (H. Chen and P. Beiersdorfer, Can. J. Phys. {\bf 86}, 55 (2008)), the authors find a significant effect due to the Breit interaction between the free and the bound electrons in the excitation process of He-like xenon. The authors state that the agreement between the measured and calculated cross-section values can only be found when the generalized Breit interaction is included in the calculations. We have performed new calculations with a Multi-Configuration Dirac-Fock code, as well as with the Penn State University suite of codes, and our conclusions are that the contribution of the Breit interaction is much lower than found in the calculations presented in the abovementioned article. In fact, our predictions are subsequently almost twice as large as the experimental values. We present these considerations in hopes of motivating new experimental investigations.
\end{abstract}


\section{Introduction}

A few years ago, in a very useful review paper entitled ``Electron-impact excitation cross-section measurements at Electron Beam Ion Traps (EBITs) from 1986 to 2006'' (H. Chen and P. Beiersdorfer, Can. J. Phys. {\bf 86}, 55 (2008) \cite{Chen08}), Chen and Beiersdorfer published values of electron-impact excitation (EIE) cross-sections for hydrogen-like Xe$^{53+}$ and helium-like Xe$^{54+}$. Such cross-sections were measured on SuperEBIT \cite{Knapp93} at electron beam energies of 112 keV and first published by Widmann {\it et al.} \cite{Widmann00}. In order to perform the measurements, new technical details had to be properly addressed. The energies of the He-like Xe$^{52+}$ K$\alpha$ excitation lines are around 30 keV, and the corresponding radiative-recombination lines at about 122 keV. This type of measurement has two difficulties:

\begin{itemize}
\item Position-sensitive proportional counters used to detect X-rays at much lower energies are no longer efficient.
\item The resolving power of crystal spectrometers in a typical Johann or von H\'amos geometry is no longer adequate due to the small Bragg angle required.
\end{itemize}

In order to overcome those difficulties, Widmann {\it et al.} used a combination of a high-purity Ge detector and a transmission-type crystal spectrometer \cite{Widmann97}. In both refs. \cite{Chen08} and \cite{Widmann00}, the authors find that the experimental EIE cross-sections derived from the Xe spectrum agree well with relativistic distorted-wave (DW) calculations as long as the Breit interaction is included. They conclude: ``This was the first experimental demonstration of this effect, and it clearly shows for the resonance line $w$ and the intercombination line $y$, where inclusion of the generalized Breit interaction changes the results by a factor of two.'' 

\section{Analysis of the results of references \cite{Chen08} and \cite{Widmann00}}

In order to compare with the experimental and calculated EIE cross-sections reported in Refs.~\cite{Chen08,Widmann00}, we (J.-C. Pain and M. Comet) performed DW computations with our Multi-Configuration Dirac-Fock (MCDF) code \cite{Bruneau83,Bruneau84,Comet15,Comet17} for the lines Ly$_{\alpha 1}$ ($1s\left(^2S_{1/2}\right)\rightarrow 2p\left(^2P_{3/2}\right)$), Ly$_{\alpha 2,3}$ ($1s\left(^2S_{1/2}\right)\rightarrow 2p\left(^2P_{1/2}\right)$, and $1s\left(^2S_{1/2}\right)\rightarrow 2s\left(^2S_{1/2}\right)$) in the case of H-like Xe$^{53+}$, and for the lines $w$ ($1s2p\left(^1P_1\right)\rightarrow 1s^2\left(^1S_0\right)$) and $y$ ($1s2p\left(^3P_1\right)\rightarrow 1s^2\left(^1S_0\right)$) in the case of He-like Xe$^{52+}$. For atoms with two or more electrons, the Breit interaction, and more recently the generalized Breit interaction, which is described by a complex function, can be included in the matrix elements of collisional processes \cite{Grant76,Grant80,Bar88,Fontes93,Zhang93,Fontes99,Sampson09}.

For high $Z$ ions, it is necessary to take into account the effects of the generalized Breit interaction, which can be derived from quantum electrodynamics (QED) via first-order perturbation theory and represents the lowest-order Feynman diagram for the exchange of a single virtual photon between two electrons. Specifically, the generalized Breit interaction, which is to be added to the Coulomb interaction expressed as $1/r_{ij}$ ($r_{ij}$ being the distance between electrons $i$ and $j$), is given by

\begin{equation}\label{gbif}
H_{GBI}(ij)=-\left(\boldsymbol{\alpha}_i.\boldsymbol{\alpha}_j\right)\frac{\exp(i\omega r_{ij})}{r_{ij}}+\left(\boldsymbol{\alpha}_i.\boldsymbol{\nabla}_i\right)\left(\boldsymbol{\alpha}_j.\boldsymbol{\nabla}_j\right)\frac{\exp(i\omega r_{ij})-1}{\omega^2r_{ij}},
\end{equation}

\noindent where $\omega$ is the frequency of the exchanged virtual photon and the $\boldsymbol{\alpha}_i$ are the usual Dirac matrices. In Eq.~(\ref{gbif}), we have used atomic units. For intermediate $Z$ values, an accurate approximation to the generalized Breit interaction can be obtained by taking the $\omega\rightarrow 0$ limit of Eq.~(\ref{gbif}), resulting in the original Breit interaction \cite{Breit29,Breit30,Breit32}:

\begin{equation}\label{gbf}
H_{B}(ij)=-\frac{1}{2r_{ij}}\left[\boldsymbol{\alpha}_i.\boldsymbol{\alpha}_j+\left(\boldsymbol{\alpha}_i.\boldsymbol{\hat{r}}_{ij}\right)\left(\boldsymbol{\alpha}_j.\hat{\boldsymbol{r}}_{ij}\right)\right],
\end{equation}

\noindent where $\hat{\boldsymbol{r}}_{ij}$ is a unit vector along $\boldsymbol{r}_{ij}$. This interaction is simply refered to as ``the Breit interaction''. It represents one of the earliest attempts to take into account the lowest-order relativistic effects associated with retardation and the magnetic interaction. The latter effect is similar to the spin-orbit interaction, but arises from the interaction between an electron and the magnetic field of another electron, rather than the field of the nucleus. Retardation is the term describing the delay in the electromagnetic interaction, which is mediated by photons, due to the finite value of the speed of light. In this case, as the electron velocities approach the speed of light, the effect of retardation becomes more important. This original form of the Breit interaction has the advantage of being computationally simpler to calculate, but its range of validity is limited due to the approximation of taking the limit $\omega\rightarrow 0$. The imaginary part of the generalized Breit interaction, Eq.~(\ref{gbif}), is important, because the impact-excitation cross-section depends on the square of the modulus of its matrix elements.

Table~\ref{Xe_CS} presents the EIE cross-sections of Xe$^{53+}$ for a scattered-electron energy of 112 keV. These comparisons show that our MCDF results are very close to the experimental values.
However, for the He-like Xe$^{52+}$ $w$ and $y$ lines, our results (see table~\ref{Xe_CS2}) are very different from the experimental values and from the calculations including the Breit interaction of Refs.~\cite{Chen08} and \cite{Widmann00}. For completeness, we note that the values including the Breit interaction (B) are very similar to those including the real-only part of the generalized Breit interaction (not shown) and to those including the complete generalized Breit interaction (GBI)\footnote{We believe that, beyond the discrepancies raised in the present paper, since Breit and generalized Breit interactions give very similar cross-sections for mid-Z elements such as Xe, the work by Widmann \emph{et al.} would have demonstrated the importance of the Breit interaction, not of the generalized Breit interaction. As far as we know, the first experimental verification of the GBI effects for EIE was made by Gumberidze \emph{et al.} \cite{Gumberidze13}. The authors studied the K-shell excitation of H-like uranium (U$^{91+}$) in relativistic collisions with different gaseous targets at the experimental storage ring at GSI (Gesellschaft f\"ur SchwerIonenforschung) in Darmstadt (Germany). Since the analysis of the results also required the inclusion of PIE (proton impact excitation), the latter experiment is not as straightforward as an EBIT measurement, but it appears to demonstrate the necessity to include the GBI in the calculations.}. Our MCDF results are consistent with calculations performed by C.J. Fontes using the Penn State University (PSU) suite of codes, which employs the DW approach as well \cite{Fontes93,Fontes99,Sampson09} (see table~\ref{Xe_CS3}).

The DW method employed in both the MCDF and PSU calculations presented here is described in detail in Section~4 of Ref.~\cite{Sampson09}. In brief, the DW method involves a partial-wave expansion of the incident and scattered electron wavefunctions. These partial waves are solutions of the Dirac equation with a distortion potential described by the charge density of the bound electrons. The excitation cross-section, or collision strength, can be expressed as a double summation of partial waves (see Eq.~(4.6) of Ref.~\cite{Sampson09}), which can be further reduced to products of angular coefficients and radial integrals involving bound and continuum electron wavefunctions. The matrix elements of the Coulomb and Breit interactions are characterized by distinctly different types of radial integrals (compare Eqs.~(4.17) and (4.18) with Eqs.~(4.78) and (4.79) of Ref.~\cite{Sampson09}) and corresponding angular coefficients. A major difference between the PSU and the MCDF codes concerns the treatment of exchange when calculating the bound electron wavefunctions. The MCDF computations use the exact, non-local exchange potential, while the PSU results employ the Dirac-Fock-Slater potential, which includes the more approximate, local Slater exchange term \cite{Slater51} with the Kohn-Sham coefficient \cite{Kohn65}. 

A second difference between the two codes is that a single mean configuration, $1s^{1.5}\, 2s^{0.16}\, 2p_{1/2}^{0.17}\, 2p_{3/2}^{0.17}$, is used to obtain both the bound and continuum wavefunctions in the PSU calculations, which ensures orthogonality among those wavefunctions. The MCDF codes use the Dirac-Fock potential in determining the bound wavefunctions and the distorted-wave potential to obtain the continuum wavefunctions. Another difference concerns the approximate treatment of the large angular momentum, partial-wave (or top-up) contribution to the excitation cross-section. The PSU calculations use the plane-wave-Born method, or Kummer transformation (see Section~4.9.3 of Ref.~\cite{Sampson09}), which has not yet been implemented in our MCDF code.

Note that Fontes \emph{et al.} have published values of collision strengths for He-like xenon ions at lower incident-electron energies (70, 300 and 1000 eV respectively) and found a similar relative effect due to the Breit interaction as reported in the present work. In Ref.~\cite{Fontes99}, the same authors provide values for high incident electron energies (up to 1000 Rydberg) and their results are also consistent with ours for the lines $w$ and $y$. In addition, they find that inclusion of the Breit interaction yields an enhancement by a factor of two only for the cross-section of the $1s2p\left(^3P_2\right)\rightarrow 1s^2\left(^1S_0\right)$ line, which is not considered here. Such behavior is the opposite of what is reported in Refs.~\cite{Chen08,Widmann00} for the lines $w$ and $y$, where the computed cross-section is almost divided by two when the GBI is included.

\begin{table}[!ht]
\caption{Collisional-excitation cross-sections (in barns) for H-like Xe$^{53+}$ at an electron energy of 112 keV. Comparison of the measurements at an electron beam ion trap (EBIT) \cite{Chen08,Widmann00} and MCDF calculations with the Coulomb (C) interaction, with the Coulomb plus Breit (C+B) interactions and with the Coulomb plus the GBI interactions (C+GBI). Ly$_{\alpha 1}$ line corresponds to $1s\left(^2S_{1/2}\right)\rightarrow 2p\left(^2P_{3/2}\right)$ and Ly$_{\alpha 2,3}$ represents the lines $1s\left(^2S_{1/2}\right)\rightarrow 2p\left(^2P_{1/2}\right)$ and $1s\left(^2S_{1/2}\right)\rightarrow 2s\left(^2S_{1/2}\right)$. The third and fourth columns contain calculations reported in Refs.~\cite{Chen08,Widmann00}.}\label{Xe_CS}
\smallskip
\begin{center}
{\small
\begin{tabular}{lcccccc}  
\hline
\noalign{\smallskip}
 Line & Experiment & Calc. \cite{Chen08,Widmann00} & Refs.~\cite{Chen08,Widmann00} & MCDF & MCDF & MCDF \\
      &            & (C) & (C+GBI) & (C) & (C+B) & (C+GBI) \\
\noalign{\smallskip}
\hline
Ly$_{\alpha 1}$ & 8.6 $\pm$ 1.5 & 8.256 & 8.109 & 8.253 & 7.318 & 8.042\\
Ly$_{\alpha 2,3}$ & 8.2 $\pm$ 3.4 & 6.541 & 6.787 & 6.567 & 6.401 & 6.717\\
\noalign{\smallskip}
\hline
\end{tabular}
}
\end{center}
\end{table}

\begin{table}[!ht]
\caption{Collisional-excitation cross-sections (in barns) for He-like Xe$^{52+}$ at an electron energy of 112 keV. Comparison of the measurements at an electron beam ion trap (EBIT) \cite{Chen08,Widmann00} and MCDF calculations with the Coulomb (C) interaction, with the Coulomb plus Breit (C+B) interactions and with the Coulomb plus the GBI interactions (C+GBI). $w$ line corresponds to $1s2p\left(^1P_1\right)\rightarrow 1s^2\left(^1S_0\right)$ and $y$ represents the line $1s2p\left(^3P_1\right)\rightarrow 1s^2\left(^1S_0\right)$. The third and fourth columns contain calculations reported in Refs.~\cite{Chen08,Widmann00}.}\label{Xe_CS2}
\smallskip
\begin{center}
{\small
\begin{tabular}{lcccccc}  
\hline
\noalign{\smallskip}
 Line & Experiment \cite{Chen08,Widmann00} & Calc. \cite{Chen08,Widmann00} & Calc. \cite{Chen08,Widmann00} & MCDF & MCDF & MCDF \\
      &            & (C) & (C+GBI) & (C) & (C+B) & (C+GBI) \\
\noalign{\smallskip}
\hline
$w$ & 7.0 $\pm$ 2.0 & 17.450 & 8.364 & 17.457 & 15.971 & 16.817\\
$y$ & 3.9 $\pm$ 1.5 & 7.313 & 3.842 & 7.767 & 7.343 & 7.616\\
\noalign{\smallskip}
\hline
\end{tabular}
}
\end{center}
\end{table}

\begin{table}[!ht]
\caption{Same as table~\ref{Xe_CS2} but for PSU computations.}\label{Xe_CS3}
\smallskip
\begin{center}
{\small
\begin{tabular}{lcccccc}  
\hline
\noalign{\smallskip}
 Line & Experiment \cite{Chen08,Widmann00} & Calc. \cite{Chen08,Widmann00} & Calc. \cite{Chen08,Widmann00} & PSU & PSU & PSU \\
      &            & (C) & (C+GBI) & (C) & (C+B) & (C+GBI) \\
\noalign{\smallskip}
\hline
$w$ & 7.0 $\pm$ 2.0 & 17.450 & 8.364 & 17.28 & 16.47 & 16.76\\
$y$ & 3.9 $\pm$ 1.5 & 7.313 & 3.842 & 7.666 & 7.608 & 7.704\\
\noalign{\smallskip}
\hline
\end{tabular}
}
\end{center}
\end{table}

\clearpage

Therefore: 

\begin{itemize}
\item We do not understand how the inclusion of the GBI can decrease the DW calculations by a factor of two. 
\item Instead, we find that the experimental values are twice lower than the calculated results.
\end{itemize}
 
\section{Conclusion}

We have performed calculations of EIE cross-sections of He-like xenon with a Multi-Configuration Dirac-Fock (MCDF) code and with the Penn State University suite of codes, and our values are much larger than the experimental ones obtained at SuperEBIT facility. We find that the impact of the Breit interaction, although definitely important, is not as significant as reported in Refs.~\cite{Chen08,Widmann00}, and is not likely to provide a satisfactory agreement with the experimental values. Our predictions disagree with the theoretical values displayed in table 4 of Ref.~\cite{Chen08} and table 3 of Ref.~\cite{Widmann00}, and are almost twice as large as the experimental values. 
Due to the importance of xenon in next-generation fusion devices, such as the National Ignition Facility (NIF) or International Thermonuclear Experimental Reactor (ITER), the EIE cross-sections considered here could play a fundamental role in the accurate modeling of relevant plasmas.
Therefore, we offer the present calculations to motivate new experimental investigations in order to corroborate (or invalidate) the published values.

\end{document}